\documentclass[njp,preprint]{revtex4}%
\usepackage{amsfonts}
\usepackage{amsmath}
\usepackage{amssymb}
\usepackage{graphicx}%
\begin{document}
\title{Entanglement and quantum teleportation under superposed gravitational fields}
\author{Yue Li$^{a,b}$}
\author{Baocheng Zhang$^{a}$}
\email{zhangbaocheng@cug.edu.cn}
\author{Li You$^{c}$}
\affiliation{$^{a}$School of Mathematics and Physics, China University of Geosciences,
Wuhan 430074, China}
\affiliation{$^{b}$State Key Laboratory of Magnetic Resonance and Atomic and Molecular
Physics, Wuhan Institute of Physics and Mathematics, Innovation Academy for
Precision Measurement Science and Technology, Chinese Academy of Sciences,
Wuhan, 430071, China}
\affiliation{$^{c}$State Key Laboratory of Low Dimensional Quantum Physics, Department of
Physics, Tsinghua University, Beijing 100084, China}
\begin{abstract}
The influence of gravitational field on entanglement of bipartite states is
investigated based on the recent idea of superposition states of gravitational
field. Different from earlier considerations, we study the case where the
gravitational field cannot be separated unitarily from the bipartite system in
the final stage of the interaction. When the different gravitational field
states are orthogonal, entanglement cannot be generated for an initial product
state. If the different gravitational field states are non-orthogonal,
entanglement can be generated and the amount of generated entanglement depends
on an overlap parameter between different gravitational field states. The
influence of gravitational field on the transfer of the state through quantum
teleportation is also studied, which might lead to an observable effect since
the quantum teleportation can be performed using macrocsopic object.

\end{abstract}
\maketitle

\section{Introduction}

Entanglement is regarded as the most nonclassical manifestation of quantum
formalism and presents the essential resources in quantum information
processing tasks (e.g. superdense coding, quantum teleportation, etc.)
\cite{hhh09}. Meanwhile, quantum entanglement is fragile and is easily
influenced by the outer environment. Classical gravity is such a general
environment and always leads to the decoherence of quantum states, which also
caused the loss of entanglement for initial entangled quantum states
\cite{bgu17}. It is unclear whether gravity as a quantum entity leads always
to the loss of entanglement. This is worthwhile to be studied since the
description of the early universe and black holes implies gravity should have
one quantum theory \cite{sc08,ew12,kw14,fd13,kst21}. For our aim to study the
influence of quantum gravitational field on entanglement, the crucial task is
how to describe the gravitational interaction using the quantum theory.

In recent years, Bose et al. \cite{sa17} and Marletto and Vedral \cite{cv17}
have proposed an effective and novel method to detect quantum properties of
gravity by considering entanglement generation through gravitational
interaction between two massive particles initially in a product state.
Following the Bose-Marletto-Vedral (BMV) proposal, extensive discussions
\cite{hr17,ah18,mv19,bwa18,mv18,cr19,nb20,mmy21,rcb21} have been carried out
and feasible methods \cite{mvd18,bcm18,cbu19,ktp20} have been proposed in
different physical systems. Although the BMV proposal has not been confirmed
experimentally, it provides an effective approach to analyze the influence of
gravitational field on entanglement between particles \cite{nb20,mmy21}. In
this work, we explore the influence of gravitational field on entanglement for
a general bipartite state further, and focus on the states of gravitational
field accompanying different components of superposed particle's states being
not completely orthogonal. We also consider the case of an initial Bell state
\cite{jsb64,nc00} of maximal entanglement, and show that gravitational field
would not affect entanglement for some situations. The Bell state will evolve,
so does the quantum process of teleportation \cite{bbw93}. In this paper, we
will also study the influence of gravitational field on quantum teleportation
and reveal a noise-like entangled quantum channel in gravitational field.

This paper is organized as follows. In the second section, we introduce the
description for the change of coherence and entanglement, by which the
influence of gravitational field on a general bipartite quantum state along
the spirit of the BMV proposal is analyzed. In the third section, we study the
situation where gravitational field cannot be separated unitarily from the
two-particle system in the final stage of their interaction. Mutual
information for estimating the change of entanglement between two particles is
then calculated for two different cases: one concerns the orthogonal states of
gravitational field, and the other when they are not orthogonal. Meanwhile,
the coherence is also discussed for every cases. Subsequently, quantum
teleportation under the influence of gravity is studied in the fourth section.
Finally, conclusions and discussions are given in the fifth section.

\section{General description}

\begin{figure}[ptb]
\centering
\includegraphics[width=5in,height=3in]{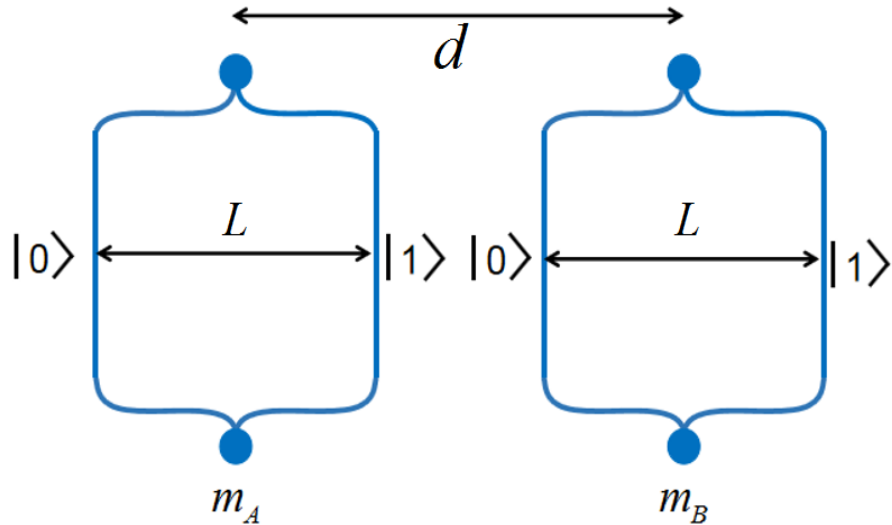} \caption{ The setting for a
two-particle system with respective masses $m_{A}$ and $m_{B}$, placed at
distance $d$ from each other. Each particle has two orthogonal eigenstates,
$|0\rangle$ and $|1\rangle$. It is required that the distance of the particle
at the state $|0\rangle$ is $L$ from that in the state $|1\rangle$,
irrespective of whether there exists entanglement between the two particles or
not. }%
\label{Fig1}%
\end{figure}

In BMW proposal, they described the quantum properties of gravitational field
using the quantum superposition of spacetime instead of giving the quantum
states for gravitational field unambiguously. For indicating the quantum
gravitational field, however, we would use the form of quantum states to sign
it, i.e. $|g\rangle$ for gravitational field in this paper. The only quantum
property is the superposition, i.e. for the same gravitational field, it could
have different quantum states, dependent on the quantum states of the source.
In particular, the BMV proposal used the first order perturbative quantum
gravity, so the interaction implemented by the gravitational field between two
massive particles can be approximated by the Newtonian form, i.e. the
interaction part of the Hamiltonian is $\frac{Gm_{A}m_{B}}{d}$ for the two
massive particles with the mass $m_{A}$, $m_{B}$, and the distance $d$ between
them. In particular, the two particles do not interact with each other
directly but through the gravitational field, which leads to the generation of
entanglement for the two particles.

When the two particles interact with the gravitational field, the whole
two-particle state would be changed and its coherence might also be changed.
It is known that the change of coherence is closely related to the change of
entanglement, but they have also the differences. The coherence derives from
the superposition of the whole quantum state and depends on the choices of
bases when measuring its change, while entanglement describes the relation
between two particles. Besides studying the change of entanglement, we expect
to discuss the change of coherence using a resource-related definition for
quantum coherence \cite{bcp14,rpl16}, $C(\rho)\equiv\min\left\Vert \rho
-\sigma\right\Vert $ where the minimization is made for the diagonal matrix
$\sigma$ in the set of incoherent states. If the norm is taken as $l_{1}%
$-norm, the measure of the coherence is obtained as
\begin{equation}
C(\rho)=\sum_{i\neq j}\left\vert \rho_{ij}\right\vert . \label{ccd}%
\end{equation}
It is shown that the $l_{1}$-norm-measure for the coherence is connected with
the success probability of unambiguous state discrimination in interference
experiments in some works \cite{mh05,bqp15}.

In order to investigate the change of entanglement between two particles
before and after the interaction of gravitational field with the particles, we
adopt mutual information \cite{nc00} as a measure. The mutual information
$I\left(  \rho_{AB}\right)  $ is a well-known physical quantity in quantum
information theory. It can measure the degree of bipartite entanglement
according to the definition,
\begin{equation}
I\left(  \rho_{AB}\right)  =S\left(  \rho_{A}\right)  +S\left(  \rho
_{B}\right)  -S\left(  \rho_{AB}\right)  , \label{mid}%
\end{equation}
where $\rho_{AB}$ is the density matrix of the two particles, $\rho
_{A}=\operatorname{Tr}_{B}\left(  \rho_{AB}\right)  $ and $\rho_{B}%
=\operatorname{Tr}_{A}\left(  \rho_{AB}\right)  $ are the reduced density
matrices of $\rho_{AB}(t)$, and $S\left(  \rho\right)  $ denotes the von
Neumann entropy $S(\rho)=-\operatorname{Tr}\left(  \rho\log_{2}\rho\right)
=-\sum_{i}\lambda_{i}\log_{2}\lambda_{i}$, given $\lambda_{i}$ as the
eigenvalues of the density matrix $\rho$. So in the following, the crucial
task is to obtain the results for the change of quantum entangled states under
the influence of gravitational field between two particles.

According to the covariant analysis of Ref. \cite{cr19}, the process of
interaction between gravitational field and particles are described
explicitly. We start with the product between a general initial state for two
particles and the gravitational field,%
\begin{equation}
|\psi_{1}\rangle=\left(  \alpha|00\rangle+\beta|01\rangle+\gamma
|10\rangle+\delta|11\rangle\right)  |g\rangle, \label{gis}%
\end{equation}
where $|\psi(0)\rangle=\alpha|00\rangle+\beta|01\rangle+\gamma|10\rangle
+\delta|11\rangle$ describes the initial state for the two particles, and the
coefficients satisfy the normalization relation, $|\alpha|^{2}+|\beta
|^{2}+|\gamma|^{2}+|\delta|^{2}=1$. The two particles of masses $m_{A}$ and
$m_{B}$, are separated by a distance $d$. Each particle has two orthogonal
states, denoted by $|0\rangle$ and $|1\rangle$, arising from its intrinsic
properties (e.g. the spin or internal state levels). When state $|\psi
(0)\rangle$ is prepared, state $|0\rangle$ is displaced from the state
$|1\rangle$ by a distance $L$, as presented in Fig. 1. An implicit assumption
behind the state (\ref{gis}) is that it has to be prepared quickly during
which cumulative interaction effect with gravity is negligible. It evolves
under the influence of gravity for a while, and the state becomes,
\begin{equation}
|\psi_{2}\rangle=\alpha e^{-i\frac{H_{00}}{\hbar}t}|00\rangle|g_{00}%
\rangle+\beta e^{-i\frac{H_{01}}{\hbar}t}|01\rangle|g_{01}\rangle+\gamma
e^{-i\frac{H_{10}}{\hbar}t}|10\rangle|g_{10}\rangle+\delta e^{-i\frac{H_{11}%
}{\hbar}t}|11\rangle|g_{11}\rangle, \label{pgs}%
\end{equation}
where the Newtonian gravitational potential is considered as an interaction
potential between the two massive particles with $H_{00}=H_{11}=-\frac
{Gm_{A}m_{B}}{d}$, $H_{01}=-\frac{Gm_{A}m_{B}}{d+L}$, $H_{10}=-\frac
{Gm_{A}m_{B}}{d-L}$ according to the proposed configuration of Fig. 1. The
two-particle state becomes entangled with gravitational field. This implies
the coherence of the particle state could be lost, although this loss of
coherence is not necessarily real for some situations, as discussed in Ref.
\cite{wgu11} where coherence can be restored by some unitary operations.
Whether the coherence of two-particle state changes depends on the
orthogonality of the states of gravitational field and will be discussed in
the next section.

Finally, the two components of each particle are brought back together, and
the position distribution for the two particles returns to one location, which
transforms all four states of gravitational field into the same one, ending up
in the state%
\begin{equation}
|\psi_{3}\rangle=(\alpha e^{-i\frac{H_{00}}{\hbar}t}|00\rangle+\beta
e^{-i\frac{H_{01}}{\hbar}t}|01\rangle+\gamma e^{-i\frac{H_{10}}{\hbar}%
t}|10\rangle+\delta e^{-i\frac{H_{11}}{\hbar}t}|11\rangle)|g\rangle.
\label{gfs}%
\end{equation}
The final state is a product state between the particles and gravitational
field, and the density operator of the particles is expressed as
\begin{equation}
\rho(t)=\left[
\begin{array}
[c]{cccc}%
|\alpha|^{2} & \alpha\beta^{\ast}e^{i\frac{\Delta_{1}}{\hbar}t} & \alpha
\gamma^{\ast}e^{i\frac{\Delta_{2}}{\hbar}t} & \alpha\delta^{\ast}\\
\alpha^{\ast}\beta e^{-i\frac{\Delta_{1}}{\hbar}t} & |\beta|^{2} & \beta
\gamma^{\ast}e^{i\frac{\Delta_{3}}{\hbar}t} & \beta\delta^{\ast}%
e^{-i\frac{\Delta_{1}}{\hbar}t}\\
\alpha^{\ast}\gamma e^{-i\frac{\Delta_{2}}{\hbar}t} & \beta^{\ast}\gamma
e^{-i\frac{\Delta_{3}}{\hbar}t} & |\gamma|^{2} & \gamma\delta^{\ast}%
e^{-i\frac{\Delta_{2}}{\hbar}t}\\
\alpha^{\ast}\delta & \beta^{\ast}\delta e^{i\frac{\Delta_{1}}{\hbar}t} &
\gamma^{\ast}\delta e^{i\frac{\Delta_{2}}{\hbar}t} & |\delta|^{2}%
\end{array}
\right]  , \label{psd}%
\end{equation}
where
\begin{align}
\Delta_{1}  &  =-\left(  H_{00}-H_{01}\right)  =Gm_{A}m_{B}\left(  \frac{1}%
{d}-\frac{1}{d+L}\right)  ,\nonumber\\
\Delta_{2}  &  =-\left(  H_{00}-H_{10}\right)  =Gm_{A}m_{B}\left(  \frac{1}%
{d}-\frac{1}{d-L}\right)  ,\nonumber\\
\Delta_{3}  &  =-\left(  H_{01}-H_{10}\right)  =Gm_{A}m_{B}\left(  \frac
{1}{d+L}-\frac{1}{d-L}\right)  .
\end{align}
\ \

As seen from the quantum states (\ref{pgs}) to (\ref{gfs}), the coherence
changes using the formula in Eq. (\ref{ccd}) which means that the unitary
operations in the disentangled process are made for the combination of quantum
states for the particles and the gravitational field, and the energy is cost
to transfer the particles to the same location in the final step. But compared
with the initial state $|\psi(0)\rangle$, the coherence of the two-particle
state does not change, i.e. $C_{i}(\rho)=C_{t}(\rho)=2(\left\vert \alpha
\beta\right\vert +\left\vert \alpha\gamma\right\vert +\left\vert \alpha
\delta\right\vert +\left\vert \beta\gamma\right\vert +\left\vert \beta
\delta\right\vert +\left\vert \gamma\delta\right\vert )$ where the superscript
$i$ ($t$) represents the coherence at the initial time (at any time $t$). This
can be checked using the initial maximal entangled state, e.g. $\alpha
=\delta=0$ and $\beta=\gamma=\frac{1}{\sqrt{2}}$, which leads to $C_{i}%
(\rho)=C_{t}(\rho)=1$. This implies that the disentangled process is indeed
realized using the unitary operations if the interaction between the particles
and gravitational field is assumed to be unitary. Since the coherence is not
changed, what about entanglement?

Using the definition Eq. (\ref{mid}), entanglement of the initial state
$|\psi(0)\rangle$ is calculated to be
\begin{equation}
I_{i}=-2\left(  \lambda_{1}\log_{2}\lambda_{1}+\lambda_{2}\log_{2}\lambda
_{2}\right)  , \label{imi}%
\end{equation}
where $\lambda_{1}$ and $\lambda_{2}$ are the eigenvalues of the density
operator $\rho_{A/B}=\operatorname{Tr}_{B/A}\left(  |\psi(0)\rangle\langle
\psi(0)|\right)  $ given by the expressions,%
\begin{align}
\lambda_{1}  &  =\frac{1}{2}+\frac{1}{2}\sqrt{1-4K_{0}K_{0}^{\ast}%
},\nonumber\\
\lambda_{2}  &  =\frac{1}{2}-\frac{1}{2}\sqrt{1-4K_{0}K_{0}^{\ast}}.
\end{align}
with $K_{0}=\alpha\delta-\beta\gamma$ and star $\ast$ denoting complex
conjugation. The initial state $|\psi(0)\rangle$ is a product state when
$K_{0}=0$ or $\alpha\delta=\beta\gamma$, its mutual information then reduces
to $I=0$. For $|\psi(0)\rangle$ a maximal entangled state, e.g., a Bell state
with $\alpha=\delta=0$ and $\beta=\gamma=\frac{1}{\sqrt{2}}$, the result
becomes $I=2$, as expected.

For the state in Eq. (\ref{psd}) after interaction with gravitational field,
its mutual information reduces to
\begin{equation}
I_{g}=-2\left(  \lambda_{3}\log_{2}\lambda_{3}+\lambda_{4}\log_{2}\lambda
_{4}\right)  \label{gmi}%
\end{equation}
where $\lambda_{3}$ and $\lambda_{4}$ are the eigenvalues of the density
operator $\rho_{A/B}=\operatorname{Tr}_{B/A}\left(  \rho_{AB}\right)  $ given
by,%
\begin{align}
\lambda_{3}  &  =\frac{1}{2}+\frac{1}{2}\sqrt{1-4K_{1}K_{1}^{\ast}%
},\nonumber\\
\lambda_{4}  &  =\frac{1}{2}-\frac{1}{2}\sqrt{1-4K_{1}K_{1}^{\ast}},
\end{align}
where $K_{1}=\left(  \alpha\delta e^{i\frac{\Delta_{1}+\Delta_{2}}{\hbar}%
t}-\beta\gamma\right)  $.

Comparing the mutual information $I_{i}$ in Eq. (\ref{imi}) with $I_{g}$ in
Eq. (\ref{gmi}), it is not hard to find that for an initial product state
specified by $K_{0}=0$, the final state would not be a product state except at
specific instants. Taking $\alpha=\beta=\gamma=\delta=\frac{1}{2}$ for a
product state, we calculate the corresponding mutual information as shown in
Fig. 2, which represents entanglement generated and confirms for gravitational
field in a quantum superposition of states, as suggested in the BMV proposal.
Entanglement is changed with time, which is different from coherence. In
particular, when the Bell state with $\alpha=\delta=0$, $\beta=\frac{1}%
{\sqrt{2}}$, and $\gamma=\pm\frac{1}{\sqrt{2}}$ is taken, the mutual
information $I_{g}=I_{i}=2$ is obtained, which means that gravitational field
does not influence entanglement between two maximally entangled particles.
Since entanglement is usually regarded as a resource in quantum information,
we will investigate in later sections on whether this invariance could hold
for operations like quantum teleportation. Moreover, other trivial situations
like $\beta=\gamma=0$, $\alpha=\frac{1}{\sqrt{2}}$, $\delta=\pm\frac{1}%
{\sqrt{2}}$ are consistent with invariance of entanglement under gravity,
which is related to the assumed form of the initial state, i.e. the distance
between the components of $|0\rangle$ is the same as that with $|1\rangle$ for
the two particles.

\begin{figure}[ptb]
\centering
\includegraphics[width=5in,height=3in]{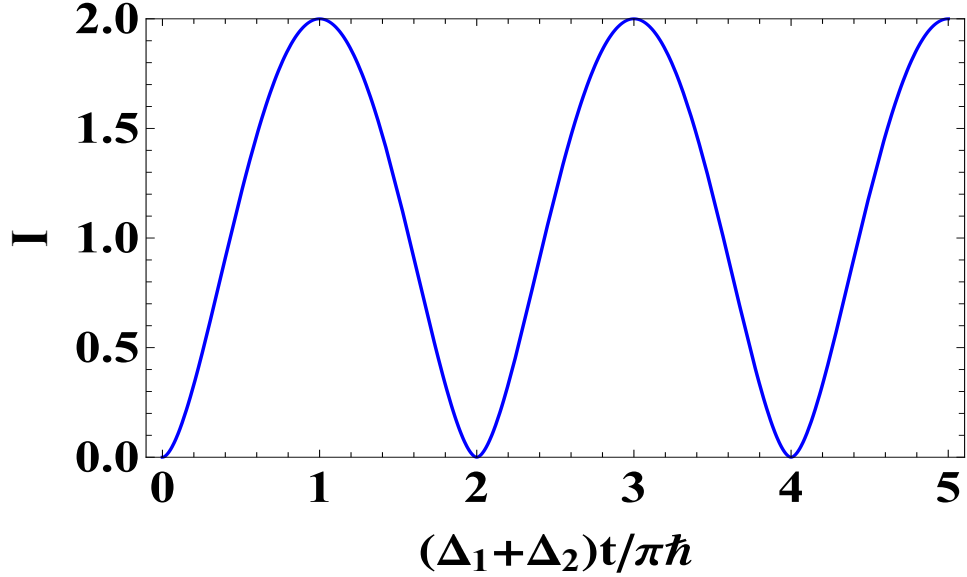} \caption{Time evolution of
mutual information under gravitational interaction when the two-particle
system is initially in a product state. }%
\label{Fig2}%
\end{figure}

\section{Influence From The Reduction}

There exist some situations where the two-particle state cannot be
disentangled from its interaction with gravitational field, even though
gravitational field is regarded as quantum. For example, with two two-level
atoms trapped in an optical cavity \cite{hmh97}, when a single photon with
proper frequency is sent into the (running wave) cavity, the state of the two
atoms would become a Bell state with the photon absorbed by either atom.
Whichever atom absorbs the photon, its position would be changed slightly due
to the backaction of photon momentum, which gives a concrete realization of
the state illustrated in Fig. 1. On the other hand, the randomness associated
with the absorption (or not absorbing) makes it impossible for the positions
to return by unitary operations for the different components of each atom, so
the different states $\left\vert g_{ij}\right\rangle $ ($i,j=0,1$) of
gravitational field cannot become the same one, leading to a real loss of
coherence for the two-particle state.

In this situation, the final state after gravitational interaction becomes
$|\psi_{2}\rangle$ in Eq. (\ref{pgs}). In order to study change of
entanglement for the two-particle state, we trace out the gravitational field
part. The resulting density operator describing the two-particle state is
given by $\rho_{AB}=\operatorname{Tr}_{g}\left(  \left\vert \psi
_{2}\right\rangle \left\langle \psi_{2}\right\vert \right)  $.

We will respectively evaluate for two cases for orthogonal or non-orthogonal
quantum states of gravitational field. First when the different quantum states
of gravitational field are orthogonal, $\left\langle g_{ij}|g_{kl}%
\right\rangle =0$, iff the subscript $ij$ is not completely same with $kl$.
The density operator describing the two-particle state reduces to
\begin{equation}
\rho_{AB}=\left[
\begin{array}
[c]{cccc}%
|\alpha|^{2} & 0 & 0 & 0\\
0 & |\beta|^{2} & 0 & 0\\
0 & 0 & |\gamma|^{2} & 0\\
0 & 0 & 0 & |\delta|^{2}%
\end{array}
\right]
\end{equation}
This is a diagonal matrix and its coherence loses completely according to Eq.
(\ref{ccd}). Its mutual information becomes%
\begin{equation}
I_{ge}=S_{A}+S_{B}-S_{AB}%
\end{equation}
where $S_{A}=S\left(  \rho_{A}\right)  =S\left(  \operatorname{Tr}_{B}\left(
\rho_{AB}\right)  \right)  $, $S_{B}=S\left(  \rho_{B}\right)  =S\left(
\operatorname{Tr}_{A}\left(  \rho_{AB}\right)  \right)  $, and $S_{AB}%
=S\left(  \rho_{AB}\right)  $ stand for von Neumann entropies for the
different density operators $\rho_{A}$, $\rho_{B}$, and $\rho_{AB}$. Their
calculated results are%
\begin{align}
S_{A}  &  =-\left(  |\alpha|^{2}+|\beta|^{2}\right)  \log_{2}\left(
|\alpha|^{2}+|\beta|^{2}\right)  -\left(  |\gamma|^{2}+|\delta|^{2}\right)
\log_{2}\left(  |\gamma|^{2}+|\delta|^{2}\right)  ,\\
S_{B}  &  =-\left(  |\alpha|^{2}+|\gamma|^{2}\right)  \log_{2}\left(
|\alpha|^{2}+|\gamma|^{2}\right)  -\left(  |\beta|^{2}+|\delta|^{2}\right)
\log_{2}\left(  |\beta|^{2}+|\delta|^{2}\right)  ,\\
S_{AB}  &  =-|\alpha|^{2}\log_{2}\left(  |\alpha|^{2}\right)  -|\beta|^{2}%
\log_{2}\left(  |\beta|^{2}\right)  -|\gamma|^{2}\log_{2}\left(  |\gamma
|^{2}\right)  -|\delta|^{2}\log_{2}\left(  |\delta|^{2}\right)  .
\end{align}
It is surprising to note despite of interaction with quantum gravitational
field, entanglement for the two-particle state does not change with time. For
an initial product state with $K_{0}=0$, entanglement cannot be generated
($I_{ge}\equiv0$) for a completely mixed final state $\rho_{AB}$. It is
interesting to note that gravitational field can be superposed although
entanglement is not generated if the states of gravitational field are
orthogonal and cannot be separated unitarily from the two-particle state.
Assuming an initial Bell state, the mutual information decreases from
$I_{i}=2$ to $I_{ge}=1$ even for the trivial case discussed in the last
section, which provides an obvious signal for the gravitational influence if
other environmentally decoherent elements \cite{whz03} are absent.

Next for the second case where different quantum states of gravitational field
are nonorthogonal. Without loss of generality, we assume that only
$\left\langle g_{01}|g_{10}\right\rangle =k\neq0$ where $k$ is a complex
nonzero number while other ones are orthogonal. The same calculation can be
extended to the more general case where none of the gravitational states are
mutually orthogonal, although the calculations and presentations become more
complex but adds no new results. The possible existence of these cases can be
understood since the distance between two components of each particles is not
large enough to make gravitational fields generated by the two components
obviously different. So it is possible to have nonzero overlaps for the
different states of gravitational field.

For this second case, tracing out gravitational field, the normalized density
matrix can be obtained as
\begin{equation}
\rho_{AB}^{\prime}=D\left[
\begin{array}
[c]{cccc}%
|\alpha|^{2} & 0 & 0 & 0\\
0 & |\beta|^{2}\left(  |k|^{2}+1\right)  & 2\beta\gamma^{\ast}e^{\frac
{\Delta_{3}t}{\hbar}}k^{\ast} & 0\\
0 & 2\beta^{\ast}\gamma e^{-i\frac{\Delta_{3}t}{\hbar}}k & |\gamma|^{2}\left(
|k|^{2}+1\right)  & 0\\
0 & 0 & 0 & |\delta|^{2}%
\end{array}
\right]  ,
\end{equation}
where $D=\frac{1}{1+|k|^{2}\left(  |\beta|^{2}+|\gamma|^{2}\right)  }$. It
shows that the coherence and entanglement for the two-particle state remains
independent of time although the state itself is time-dependent. The general
expression for mutual information becomes a little complicated to give out in
this case, so instead we will discuss several specific situations.

For a direct product state, $K_{0}=0$, we choose $\alpha=\beta=\gamma
=\delta=\frac{1}{2}$, and the result of mutual information can be expressed
as
\begin{align}
I_{gd}=  &  2+2\left(  \frac{1}{4+2|k|^{2}}\right)  \log_{2}\left(  \frac
{1}{4+2|k|^{2}}\right)  +\left(  \frac{(|k|+1)^{2}}{\left[  4+2|k|^{2}\right]
}\right)  \log_{2}\left(  \frac{(|k|+1)^{2}}{\left[  4+2|k|^{2}\right]
}\right) \nonumber\\
&  +\left(  \frac{(|k|-1)^{2}}{\left[  4+2|k|^{2}\right]  }\right)  \log
_{2}\left(  \frac{(|k|-1)^{2}}{\left[  4+2|k|^{2}\right]  }\right)  ,
\end{align}
which depends on $k$ but is independent of time. If $k$ is related to the
difference between different gravitational fields, with $k=0$ denoting the
maximal difference as presented in first case where states of gravitational
field are mutually orthogonal, and $k=1$ for the minimal difference as
discussed in the last section when the states of gravitational field are
separable from the two-particle system. The dependence on parameter $k$ is
illustrated in the upper plot of Fig. 3. It is seen that the mutual
information is monotonically increasing with parameter $k$, which is
consistent with the change of coherence calculated by Eq. (\ref{ccd}),
$C=\frac{2k}{2+|k|^{2}}$ and $\frac{dC}{dk}=\frac{4-2|k|^{2}}{\left(
2+|k|^{2}\right)  ^{2}}>0$ for $0\leq\left\vert k\right\vert \leq1$.

\begin{figure}[ptb]
\centering
\includegraphics[width=5in,height=3in]{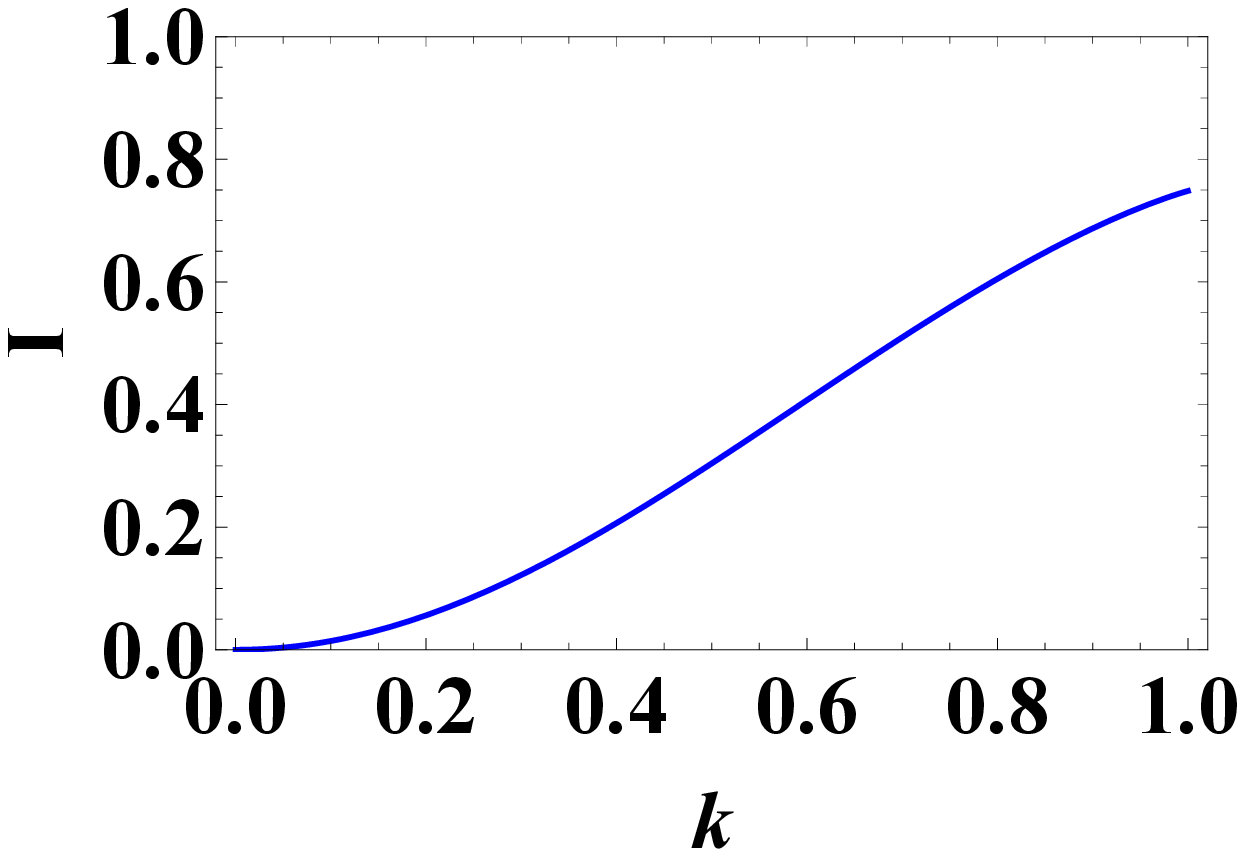}\newline%
\includegraphics[width=5in,height=3in]{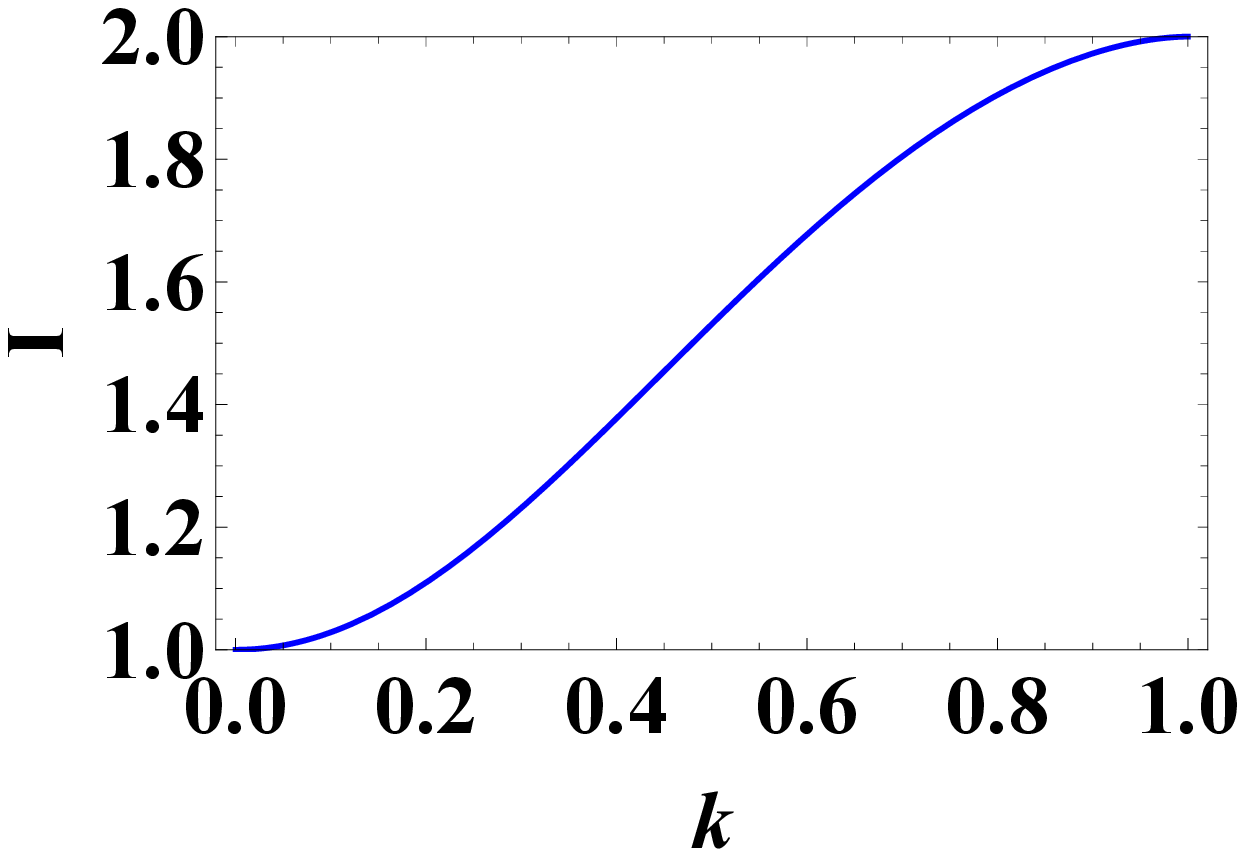}\newline\caption{Mutual
information as a function of parameter $k$ when gravitational field states are
non orthogonal. The upper plot describes the case of an initial two- particle
product state with $\alpha=\beta=\gamma=\delta=\frac{1}{2}$. The lower plot
describes the case of an initial two-particle Bell state with $\alpha
=\delta=0$ and $\beta=\gamma=\frac{1}{\sqrt{2}}$. }%
\label{Fig3}%
\end{figure}

For Bell states, two different situations arise. For the trivial situation
discussed in the last section, $\beta=\gamma=0$, $\alpha=\frac{1}{\sqrt{2}}$,
and $\delta=\pm\frac{1}{\sqrt{2}}$, the mutual information is independent of
$k$ and is equal to $1$ as in the first case. However, for situations with
$\alpha=\delta=0$, $\beta=\frac{1}{\sqrt{2}}$, and $\gamma=\pm\frac{1}%
{\sqrt{2}}$, the mutual information becomes
\begin{equation}
I_{gb}=2+\left(  \frac{1}{2}+\frac{|k|}{1+|k|^{2}}\right)  \log_{2}\left(
\frac{1}{2}+\frac{|k|}{1+|k|^{2}}\right)  +\left(  \frac{1}{2}-\frac
{|k|}{1+|k|^{2}}\right)  \log_{2}\left(  \frac{1}{2}-\frac{|k|}{1+|k|^{2}%
}\right)  ,
\end{equation}
which is shown in the lower plot of Fig. 3. When $k=0$, $I_{gb}=1$, as for the
first case where gravitational field states are mutually orthogonal, and when
$k=1$, $I_{gb}=2$, as obtained in the last section where entanglement is
unchanged after interaction with gravitational field. According to the formula
(\ref{ccd}), the coherence is obtained as $C=\frac{2k}{1+|k|^{2}}$ which is
monotonically increasing with parameter $k$, i.e.$\frac{dC}{dk}=\frac
{2-2|k|^{2}}{\left(  1+|k|^{2}\right)  ^{2}}>0$ for $0\leq\left\vert
k\right\vert \leq1$, consistent with the change of entanglement presented in
the lower plot of Fig. 3.

An interesting phenomena has to be pointed out that the change trends for the
coherence and entanglement are the same for these different cases in this
section. This distinguishes from the situation that the two-particle state is
disentangled from the interaction with gravitational field, where the
coherence remains unchanged but entanglement changes with time for an
initially quantum entangled state.

\section{Influence on Quantum Teleportation}

The influence of quantum gravitational field on entanglement of the
two-particle system has been analyzed above, and it is found that there exist
some situations where entanglement is invariant under gravity. In this
section, we will investigate how gravitational field affect another quantum
entanglement process according to the BMV proposal. Specifically, we consider
the process of quantum teleportation, which describes a protocol for
transferring an unknown quantum state from one location to another
\cite{bbw93}. The two parties are commonly referred to as Alice and Bob, who
share an entangled state such as a Bell state to start with. The general
process is shown in Fig. 4.

\begin{figure}[ptb]
\centering
\includegraphics[width=5in,height=3in]{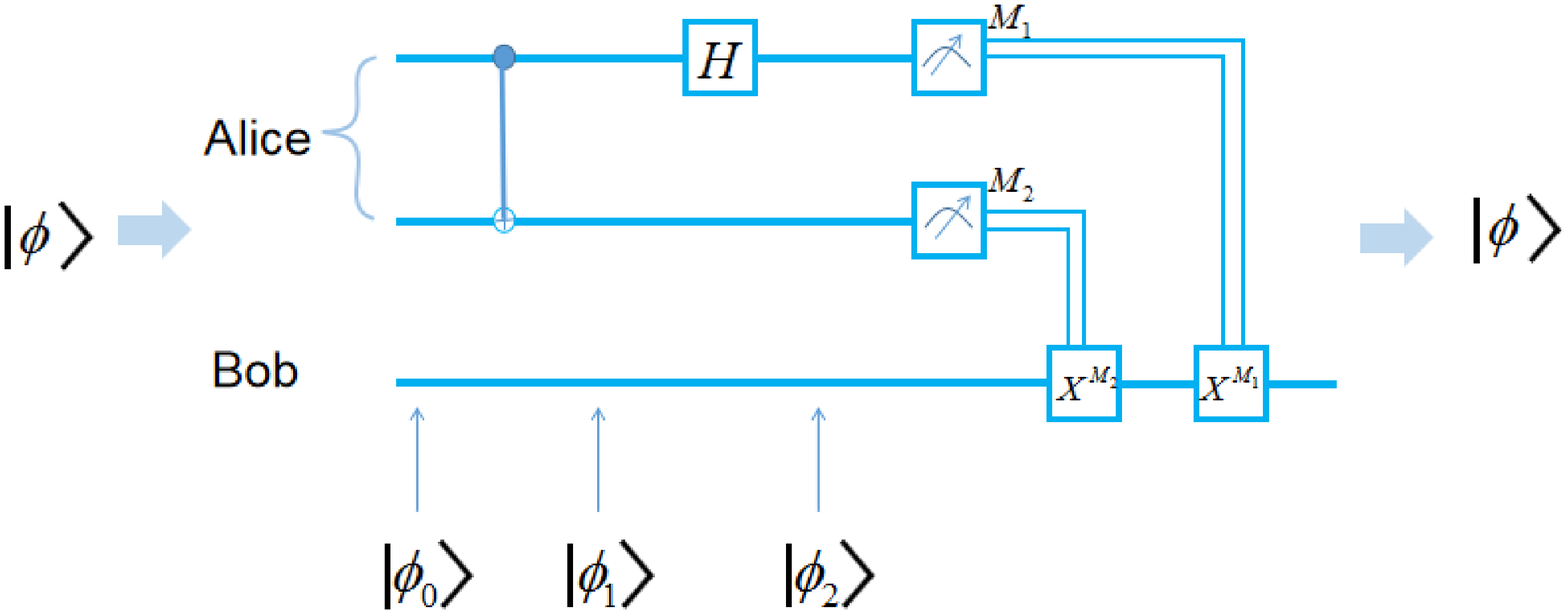} \caption{The process of
quantum teleportation \cite{nc00}. Tow top lines represent Alice's system,
while the bottom line represents Bob's system. $|\phi\rangle$ is an unknown
quantum state. After a series of operations including a CNOT gate, a Hadamard
gate (H), quantum measurement (M) and other processes (X), the state
$|\phi\rangle$ is transferred from Alice to Bob. For a more detailed
description please refer to Ref. \cite{nc00}. }%
\label{Fig4}%
\end{figure}

Suppose the teleportation happens between Alice and Bob and they share a Bell
state $\frac{1}{\sqrt{2}}(|0\rangle_{A}|1\rangle_{B}+|1\rangle_{A}%
|0\rangle_{B})$. According to the description in the above, the Bell state
becomes $|\psi_{B}\rangle=\frac{1}{\sqrt{2}}\left(  e^{-i\frac{H_{01}}{\hbar
}t}|01\rangle+e^{-i\frac{H_{10}}{\hbar}t}|10\rangle\right)  $ under the
influence of gravity. Now, an unknown quantum state
\begin{equation}
|\phi\rangle=\tilde{\alpha}|0\rangle+\tilde{\beta}|1\rangle, \label{uqs}%
\end{equation}
with $|\tilde{\alpha}|^{2}+|\tilde{\beta}|^{2}=1$, is to be transferred from
Alice to Bob. The combined state becomes initially
\begin{equation}
\left\vert \phi_{0}\right\rangle =|\phi\rangle|\psi_{B}\rangle=\frac{1}%
{\sqrt{2}}(\tilde{\alpha}|0\rangle+\tilde{\beta}|1\rangle)\left(
e^{-i\frac{H_{01}}{\hbar}}|0\rangle|1\rangle+e^{-i\frac{H_{10}t}{\hbar}%
}|1\rangle|0\rangle\right)  .
\end{equation}
After Alice implements a CNOT gate on the unknown state and her part of the
Bell state, the whole state becomes
\begin{equation}
\left\vert \phi_{1}\right\rangle =\frac{1}{\sqrt{2}}(\tilde{\alpha}%
e^{-i\frac{H_{01}}{\hbar}t}|001\rangle+\tilde{\alpha}e^{-i\frac{H_{10}}{\hbar
}t}|010\rangle+\tilde{\beta}e^{-i\frac{H_{01}}{\hbar}t}|111\rangle
+\tilde{\beta}e^{-i\frac{H_{10}}{\hbar}t}|100\rangle).
\end{equation}
Then, she sends the first qubit through a Hadamard gate, and the state evolves
into
\begin{equation}
\left\vert \phi_{2}\right\rangle =\frac{1}{2}\left(  |00\rangle|\phi
\rangle_{00}+|01\rangle|\phi\rangle_{01}+|10\rangle|\phi\rangle_{10}%
+|11\rangle|\phi\rangle_{11}\right)  , \label{hmw}%
\end{equation}
with%
\begin{align}
|\phi\rangle_{00}  &  =\tilde{\alpha}e^{-i\frac{H_{01}}{\hbar}t}%
|1\rangle+\tilde{\beta}e^{-i\frac{H_{10}}{\hbar}t}|0\rangle,\\
|\phi\rangle_{01}  &  =\tilde{\alpha}e^{-i\frac{H_{10}}{\hbar}t}%
|0\rangle+\tilde{\beta}e^{-i\frac{H_{01}}{\hbar}t}|1\rangle,\\
|\phi\rangle_{10}  &  =\tilde{\alpha}e^{-i\frac{H_{01}}{\hbar}}|1\rangle
-\tilde{\beta}e^{-i\frac{H_{10}}{\hbar}t}|0\rangle,\\
|\phi\rangle_{11}  &  =\tilde{\alpha}e^{-i\frac{H_{10}}{\hbar}t}%
|0\rangle-\tilde{\beta}e^{-i\frac{H_{01}}{\hbar}t}|1\rangle.
\end{align}

If Alice measures the first and second qubits, she will find four different
outcomes: $00$, $01$, $10$, and $11$. The corresponding quantum states
obtained by Bob would be $\left\vert \phi\right\rangle _{00}$, $\left\vert
\phi\right\rangle _{01}$, $\left\vert \phi\right\rangle _{10}$, and
$\left\vert \phi\right\rangle _{11}$. After the measurement, Alice sends the
measured results to Bob through a classical channel. Bob converts his quantum
state using the associated unitary operations according to Alice's message,
and finally obtains the same quantum state as if there were no influence of
gravity. In order to study the influence of gravity, we assume that the
measured result by Alice is $00$, and thus the Bob would obtain a state with
the form $\tilde{\alpha}e^{-i\frac{H_{01}}{\hbar}t}|0\rangle+\tilde{\beta
}e^{-i\frac{H_{10}}{\hbar}t}|1\rangle$ after an operation of X gate. After
absorbing the factor of $e^{-i\frac{H_{10}}{\hbar}t}$, the final quantum state
obtained by Bob can be reduced to
\begin{equation}
\left\vert \phi^{\prime}\right\rangle =\tilde{\alpha}|0\rangle+\tilde{\beta
}e^{i\frac{\Delta_{3}}{\hbar}t}|1\rangle. \label{fts}%
\end{equation}
Obviously, the state $\left\vert \phi^{\prime}\right\rangle $ includes a
gravitational field dependent phase and is different from the unknown quantum
state $|\phi\rangle$. In principle, one can carry out unitary operations to
transform the state $\left\vert \phi^{\prime}\right\rangle $ to $|\phi\rangle
$, but this is difficult because the influence of gravity cannot be controlled precisely.

Now we estimate the faithfulness or fidelity of the teleportation, which can
be measured by the mean distance \cite{sp94,ng96,hh96} between the transferred
state $|\phi\rangle$ and the received state $\left\vert \phi^{\prime
}\right\rangle $, $\overline{F}=\int_{S}dM\left(  \phi\right)  \sum_{k}%
p_{k}Tr(\rho^{\prime}\rho)$ where $\rho$ is the density matrix of state
$|\phi\rangle$, $\rho^{\prime}$ is the density matrix of state $\left\vert
\phi^{\prime}\right\rangle $, $p_{k}$ is the probability of obtaining the
result $k$ by Alice, and $M\left(  \phi\right)  $ denotes uniform distribution
on the Bloch sphere $S$. It has been shown \cite{sp94,ng96,hh96} that
$\overline{F}=\frac{1}{2}$ if the received state is completely independent of
the transferred state; $\overline{F}=\frac{2}{3}$ for the purely classical
channel or the channel with the product state; $\overline{F}\simeq0.87$ for
the boundary for the local and non-local states in the sense of local hidden
variables, i.e. the state of the channel reveals \textquotedblleft
nonclassical aspects\textquotedblright\ incompatible with local hidden
variables only if $\overline{F}>0.87$.

For the transferred state (\ref{uqs}), we take $\tilde{\alpha}=\cos\left(
\frac{\theta}{2}\right)  $, $\tilde{\beta}=\sin\left(  \frac{\theta}%
{2}\right)  e^{i\varphi}$, where $\theta\in\lbrack0,\pi]$ and $\varphi
\in\lbrack0,2\pi]$, and obtain the averaged fidelity as%
\begin{equation}
\overline{F}=\frac{1}{4\pi}\int_{0}^{2\pi}d\varphi\int_{0}^{\pi}F\sin\theta
d\theta, \label{afv}%
\end{equation}
where $F=\sum_{k}p_{k}Tr(\rho^{\prime}\rho)$ measures how different the
received state $\rho^{\prime}$ is from $\rho$, averaged over the probabilities
of all measured outcomes by Alice. If the transferred state and the received
state are pure, the fidelity $F\left(  |\phi\rangle,\left\vert \phi^{\prime
}\right\rangle \right)  =4\times\frac{1}{4}$ $\left\vert \left\langle
\phi|\phi^{\prime}\right\rangle \right\vert ^{2}=1-\frac{1}{2}\sin^{2}%
\theta\left[  1-\cos\left(  \frac{\Delta_{3}}{\hbar}t\right)  \right]  $ where
Alice measures every results with the same probability $\frac{1}{4}$ as seen
in Eq. (\ref{hmw}) and the fidelities for every results are also the same.
Using Eq. (\ref{afv}), we obtain%
\begin{equation}
\overline{F}\left(  |\phi\rangle,\left\vert \phi^{\prime}\right\rangle
\right)  =\frac{2}{3}+\frac{1}{3}\cos\left(  \frac{\Delta_{3}}{\hbar}t\right)
.
\end{equation}
This averaged fidelity is time-dependent and depends on gravitational field,
as shown in the upper plot of Fig. 5. Note that the maximal value of the
averaged fidelity is $1$, which shows that at some instants the teleportation
is perfect; the minimal value of the averaged fidelity is $\frac{1}{3}$, which
is hardly possible to teleport any unknown state. In the upper plot of Fig. 5,
the black line represents $\overline{F}\left(  |\phi\rangle,\left\vert
\phi^{\prime}\right\rangle \right)  =\frac{2}{3}$, above which non-product
states exist; the red line represents $\overline{F}\left(  |\phi
\rangle,\left\vert \phi^{\prime}\right\rangle \right)  \simeq0.87$, above
which the non-local states exist, unexplainable by the local hidden variables.
In particular, the limit for the averaged fidelity over time becomes
\begin{equation}
\underset{t\rightarrow\infty}{lim}\frac{\int_{0}^{t}\overline{F}dt}{t}%
=\frac{2}{3},
\end{equation}
which shows that the coherence and entanglement for the two-particle state
disappear after long enough time.

\begin{figure}[ptb]
\centering
\includegraphics[width=4.5in,height=2.8in]{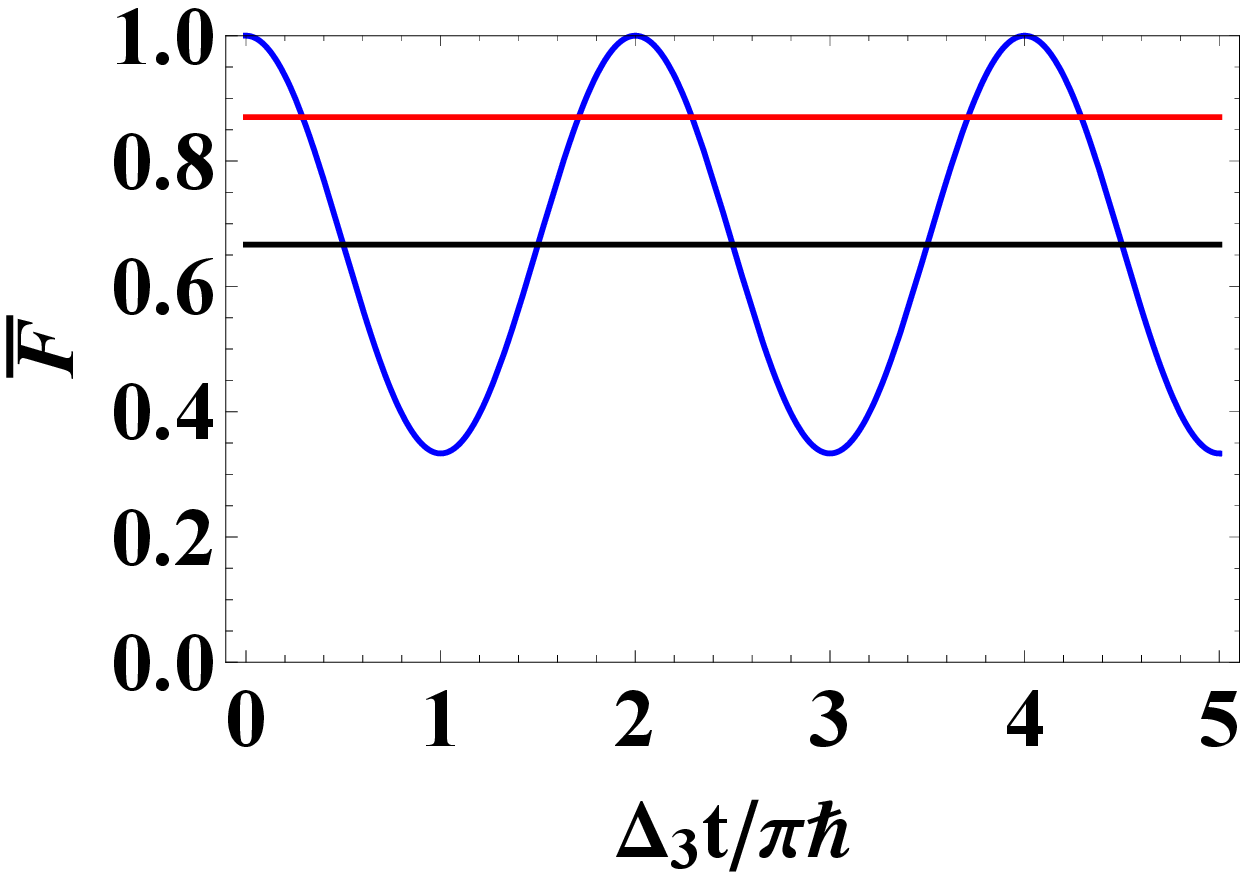}\newline%
\includegraphics[width=4.5in,height=2.8in]{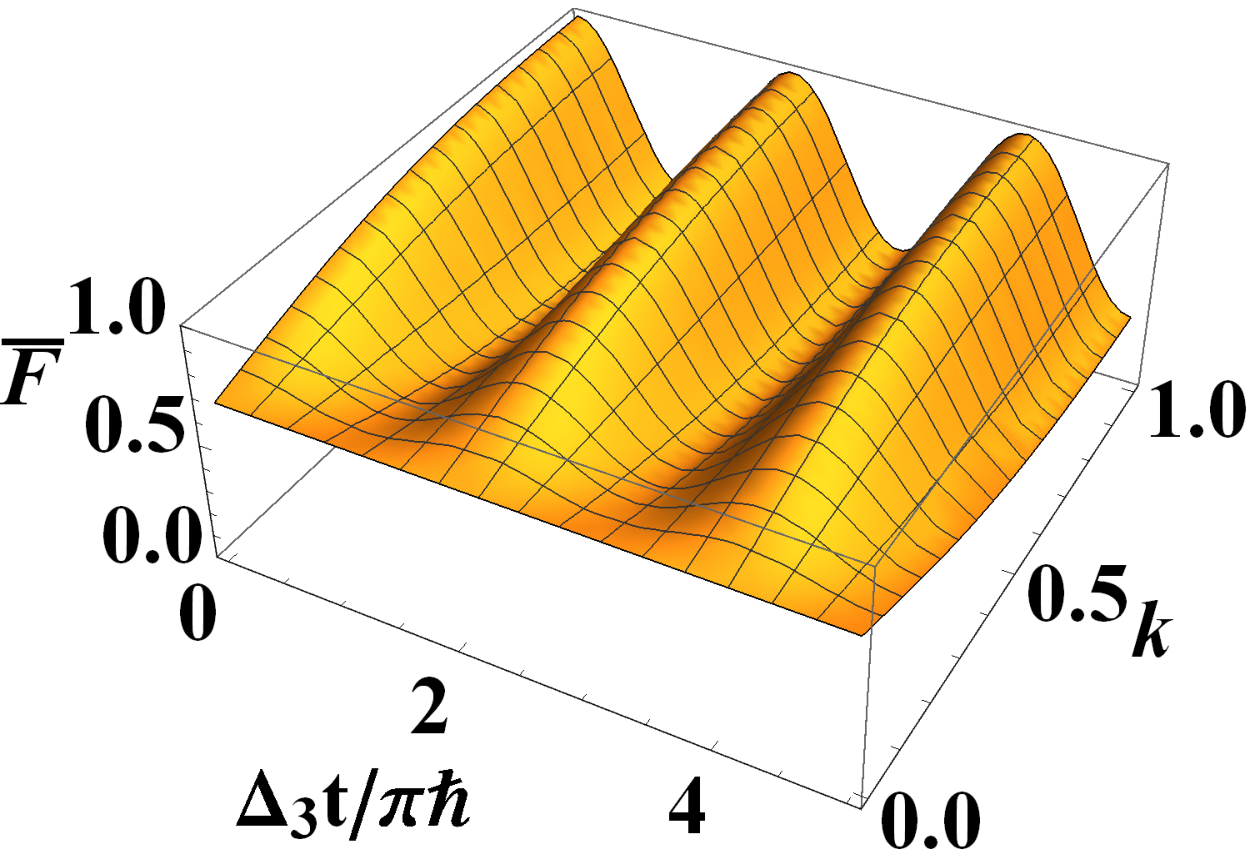}\newline\caption{ The
averaged fidelity of an unknown quantum state transferred through quantum
teleportation under the influence of gravitational field. The upper plot
represents the change of averaged fidelity with time when entanglement is not
changed by gravitational interaction. The lower plot represents the change of
averaged fidelity with time and parameter $k$ when the gravitational field
states are non-orthogonal.}%
\label{Fig5}%
\end{figure}

In general, the gravitational field states cannot be disentangled unitarily in
the final stage of the interaction, since during quantum teleportation the
required operations are difficult to implement. The Bell state as an
entanglement channel becomes
\begin{equation}
\left\vert \psi^{\prime}\right\rangle =\frac{1}{\sqrt{2}}\left(
e^{-i\frac{H_{01}}{\hbar}t}|01\rangle\left\vert g_{01}\right\rangle
+e^{-i\frac{H_{10}}{\hbar}t}|10\rangle\left\vert g_{10}\right\rangle \right)
.
\end{equation}
The teleportation as implemented between Alice and Bob goes through the same
procedures above for without gravity interaction. Before Alice performs the
measurement, the state becomes
\begin{equation}
\left\vert \phi_{2}^{\prime}\right\rangle =\frac{1}{2}\left(  |00\rangle
\left\vert \phi^{\prime}\right\rangle _{00}+|01\rangle\left\vert \phi^{\prime
}\right\rangle _{01}+|10\rangle\left\vert \phi^{\prime}\right\rangle
_{10}+|11\rangle\left\vert \phi^{\prime}\right\rangle _{11}\right)  ,
\end{equation}
where%
\begin{align}
\left\vert \phi^{\prime}\right\rangle _{00}  &  =\tilde{\alpha}e^{-i\frac
{H_{01}}{\hbar}t}|1\rangle\left\vert g_{01}\right\rangle +\tilde{\beta
}e^{-i\frac{H_{10}}{\hbar}t}|0\rangle\left\vert g_{10}\right\rangle ,\\
\left\vert \phi^{\prime}\right\rangle _{01}  &  =\tilde{\alpha}e^{-i\frac
{H_{10}}{\hbar}t}|0\rangle\left\vert g_{10}\right\rangle +\tilde{\beta
}e^{-i\frac{H_{01}}{\hbar}t}|1\rangle\left\vert g_{01}\right\rangle ,\\
\left\vert \phi^{\prime}\right\rangle _{10}  &  =\tilde{\alpha}e^{-i\frac
{H_{01}}{\hbar}t}|1\rangle\left\vert g_{01}\right\rangle -\tilde{\beta
}e^{-i\frac{H_{10}}{\hbar}t}|0\rangle\left\vert g_{10}\right\rangle ,\\
\left\vert \phi^{\prime}\right\rangle _{11}  &  =\tilde{\alpha}e^{-i\frac
{H_{10}}{\hbar}t}|0\rangle\left\vert g_{10}\right\rangle -\tilde{\beta
}e^{-i\frac{H_{01}}{\hbar}t}|1\rangle\left\vert g_{01}\right\rangle ,
\end{align}

Analogously, we suppose the measured result by Alice is $00$, so the final
state obtained by Bob becomes
\begin{equation}
\left\vert \phi^{\prime}\right\rangle =\tilde{\alpha}e^{-i\frac{H_{10}}{\hbar
}t}|0\rangle\left\vert g_{10}\right\rangle +\tilde{\beta}e^{-i\frac{H_{01}%
}{\hbar}t}|1\rangle\left\vert g_{01}\right\rangle ,
\end{equation}
which depends on gravitational field. After tracing out gravitational field,
the state obtained by Bob is described by the density operator
\begin{equation}
\rho^{\prime}=|\tilde{\alpha}|^{2}|0\rangle\langle0|+\frac{2k\tilde{\alpha
}\tilde{\beta}^{\ast}}{|k|^{2}+1}e^{-i\frac{\Delta_{3}t}{\hbar}}%
|0\rangle\langle1|+\frac{2k^{\ast}\tilde{\alpha}^{\ast}\tilde{\beta}}%
{|k|^{2}+1}e^{i\frac{\Delta_{3}}{\hbar}t}|1\rangle\langle0|+|\tilde{\beta
}|^{2}|1\rangle\langle1|,
\end{equation}
assuming $\left\langle g_{01}|g_{10}\right\rangle =k$.

The fidelity can be calculated as $F\left(  |\phi\rangle,\rho^{\prime}\right)
=4\times\frac{1}{4}\langle\phi|\rho^{\prime}|\phi\rangle=1-\frac{1}{2}%
Q\sin^{2}\theta$ with $Q=1-\frac{k^{\ast}+k}{|k|^{2}+1}\cos(\frac{\Delta_{3}%
}{\hbar}t)+i\frac{k^{\ast}-k}{|k|^{2}+1}\sin(\frac{\Delta_{3}}{\hbar}t)$.
According to the definition of averaged fidelity (\ref{afv}), we obtained
\begin{equation}
\overline{F}\left(  |\phi\rangle,\rho^{\prime}\right)  =1-\frac{1}{3}Q.
\end{equation}
When the parameter $k$ is taken as real, the fidelity takes the form
$\overline{F}\left(  |\phi\rangle,\rho^{\prime}\right)  =\frac{2}{3}%
(1+\frac{k}{|k|^{2}+1}\cos(\frac{\Delta_{3}}{\hbar}t))$, which is presented in
the lower plot of Fig. 5. It is seen from the figure that the fidelity of
transferring an unknown state is closely dependent on the properties of
gravitational field. When $k$ is large, the oscillation of the averaged
fidelity is obvious with time, which includes the states for the channel with
their averaged fidelity below $\frac{2}{3}$, and some other states that
presented the non-locality above $0.87$. When $k$ decreases, the fluctuation
becomes smaller and smaller. It is equal to $\frac{2}{3}$ for $k=0$, which
shows that the state for the channel becomes a complete mixed state or a
classical state, as discussed at the beginning of the second section, and such
states have no coherence.

\section{Conclusion and Discussion}

In this work, we investigate the influence of gravitational field on quantum
entanglement following the BMV proposal but for a general two-particle state.
It is found again that entanglement can be generated for an initial product
state inside gravitational field but the coherence is unchanged in the
process. This entanglement can be invariant with time when the initial state
is a Bell state. Both require a crucial operation to separate unitarily
gravitational field from the two-particle state in the final stage of evolution.

In some cases, gravitational field is not easily or is impossible to be
separated unitarily from the particles. We have analyzed the corresponding
processes of interaction between a two-particle system and gravitational
field. When the gravitational field states are orthogonal, the influence on
entanglement for the two-particle state is time-independent. For an initial
product state, entanglement cannot be generated through interaction with
gravitational field, since the final two-particle state is a maximally mixed
state. For a Bell state, its entanglement decreases as the mutual information
is decreased from $2$ to $1$. This is an evident signal for experimental
observation if other environmentally decoherent elements can be avoided. When
the gravitational field states are not completely orthogonal, the influence of
gravitational field on entanglement for the two-particle state depends on the
degree of overlap (measured by the parameter $k$) among the different
gravitational field states, for all the above cases considered with a tunable
parameter $k$. Whether the states for gravitational field are orthogonal or
under what conditions they could be orthogonal is unclear for the time being,
it is therefore necessary and significant to analyze the case where the
gravitational field states are non-orthogonal. Our results indeed present
different behaviors which are helpful for possible future observations. In
particular, the change of coherence has the same trend of increasing
monotonically with the change of entanglement in this case that the
two-particle state cannot be disentangled from gravitational field. While the
two-particle state is disentangled from gravitational field, the coherence
remains unchanged but entanglement presents a oscillating behaviors with time.

We have also discussed the influence of gravitational field on quantum
teleportation, and we find that an exact transfer of an unknown quantum state
is hardly possible and the relative phase containing information about
gravitational field always appears in the transferred state. In particular,
quantum teleportation can be broken for some evolved states since they have
lost the non-locality and even lost all entanglement and coherence.

Finally, we point out that the gravitational field effect discussed is very
small for microscopic particles, but could be observable in the future for
mesoscopic particles as analyzed in the initial BMV proposals. Since quantum
teleportation can be performed using entangled channel between mesoscopic or
even macroscopic physical systems \cite{hhd16,lhx20,fhg21}, the influence of
gravitational field might be observed in these experiments in the near future.

\section{Acknowledgement}

This work is supported by Grant No. 11654001 of the National Natural Science
Foundation of China (NSFC).

\end{document}